\documentclass[conference]{IEEEtran}
\usepackage{latexsym}   
\usepackage{amsfonts}
\usepackage[cmex10]{amsmath}
\usepackage{amssymb}
\usepackage{url}
\usepackage{textcomp}
\usepackage{graphicx}
\usepackage{epstopdf}
\usepackage{epsfig}     
\usepackage{rotating}       
\usepackage{eufrak}

\usepackage{algorithm}
\usepackage{algpseudocode}
\usepackage[fleqn,tbtags]{mathtools}
\usepackage{mhsetup}
\usepackage{booktabs}

\usepackage{caption}
\usepackage{subcaption}
\usepackage{float}
\usepackage[caption = false]{subfig}
\usepackage{csquotes}
\usepackage{multirow}

\ifCLASSINFOpdf
\else
\fi

\begin{document}

\title{Inferring disease causing genes and their pathways: A mathematical perspective}

\author{\IEEEauthorblockN{Jeethu V. Devasia}
\IEEEauthorblockA{Department
of Computer Science and Engineering\\
National Institute of Technology Calicut,\\
India - 673 601\\
Email: jeethu\_p130021cs @ nitc.ac.in}
\and
\IEEEauthorblockN{Priya Chandran}
\IEEEauthorblockA{Department
of Computer Science and Engineering,\\
National Institute of Technology Calicut,\\
India - 673 601\\
Email: priya  @ nitc.ac.in}}


\maketitle

\begin{abstract}
\textbf{Background and Objective:}
A system level view of cellular processes for human and several organisms can be captured by analyzing
molecular interaction networks. A molecular interaction network formed of differentially expressed genes and their interactions helps to understand key players behind disease development. So, if the functions of these genes are blocked by altering their interactions, it would have a great impact in controlling the disease. Due to this promising consequence, the problem of \textit{inferring disease causing genes and their pathways} has attained a crucial position in computational biology research. However, considering the huge size of interaction networks, executing computations can be costly. Review of literatures shows that the methods proposed for finding the set of disease causing genes could be assessed in terms of their accuracy which a perfect algorithm would find. Along with accuracy, the time complexity of the method is also important, as high time complexities would limit the number of pathways that could be found within a pragmatic time interval.\\
\textbf{Methods and Results:}
Here, the problem  has been tackled by integrating graph theoretical approaches with an approximation algorithm. The problem of \textit{inferring disease causing genes and their pathways} has been transformed to a graph theoretical problem. Graph pruning techniques have been applied to get the results in practical time. Then, randomized rounding, an efficient approach to design an approximation algorithm, has been applied to fetch the most relevant causal genes and pathways. Experimentation on multiple benchmark datasets has been demonstrated more accurate and computationally time efficient results than existing algorithms. Also, biological relevance of these results has been analyzed.\\
\textbf{Conclusions:}
Based on computational approaches on biological data, the sets of disease causing genes and corresponding pathways are identified for multiple disease cases. The proposed approach would have a remarkable contribution in areas like drug development and gene therapy, if we could recognize these results biologically too.
\end{abstract}

\begin{IEEEkeywords}
Molecular interaction Network; Causal genes; Dysregulated pathway; Graph pruning; Approximation algorithm; Randomized rounding
\end{IEEEkeywords}

\IEEEpeerreviewmaketitle

\section{Introduction}\label{intro}
Cellular processes are mainly governed by the co-action of biomolecules. For example, a particular protein function can be understood by mapping protein-protein interactions. These biological communications can be represented using molecular interaction networks. At present, molecular interaction networks of various organisms are available \cite{ref:yoo}. We can exploit them for diverse aims such as discovering disease causing genes and related pathways. From the computational perspective, computing information flows in a complex model is expensive as the input sizes are large and the analyses typically have very high time complexities. Therefore, proposing algorithms for effective computation on the networks would have a remarkable impact on the knowledge to be gained from such networks. The problem of \textit{inferring disease causing genes and dysregulated pathways} is of prime importance and huge academic and industrial interest, because, it is potentially very useful for comprehending the underlying 
system of complex diseases and suggesting prospective drug targets. An algorithm that augments graph theoretical approaches with approximation for \textit{inferring causal genes and dysregulated pathways} is addressed in this paper. The proposed method incorporates gene expression value because of its potentiality in predicting diseases. High-risk genes are more correlated with each other than the genes with lower risk and vice versa \cite{ref:win}. An experimental analysis of the state of the art related works together with the proposed method is also given in this paper. Related works are based on Random walk based approach, Electric circuit model with Expression Quantitative Loci (eQTL) analysis, Electric circuit model with multiple sources and sinks, Fast iterative matrix inversion and Approximation algorithm based on Randomized rounding \cite{ref:cho}, \cite{ref:zhi}, \cite{ref:doy}, \cite{ref:sut}, \cite{ref:yoo}, \cite{ref:yoo2}, \cite{ref:tij}, \cite{ref:jee}.\\

Proposed by Tu et al., the random walk approach has shown a significant impact in the problem of identifying \textit{causal genes} and and the underlying \textit{pathways} \cite{ref:yoo}, \cite{ref:cho}, \cite{ref:zhi}, \cite{ref:sut}. Based on the Pearson's correlation coefficient of \textit{gene expression values} of genes, transition probability is defined. Starting from a source node, random walker moves to a node that is qualified as an unvisited and highest transition probability bearing node among all the neighbors of the current node. This process is repeated until it visits the destination node or further movement is not possible. \textit{Candidate causal gene} that has the largest number of visit times or that has the largest value of probability of being a \textit{causal gene} is taken as the \textit{causal gene}, $g_c$. Identification of \textit{pathway} is done by tracing a path from $g_c$ to the corresponding source node by selecting intermediate nodes as the most visited ones. According to 
Suthram et al., this approach results in relatively short walks with the requirement of multiple iterations for better results \cite{ref:sut}. They proposed a new approach based on Electric circuit model which is analogous to Random walk based approach \cite{ref:sut}, \cite{ref:yoo}, \cite{ref:doy}, \cite{ref:cho}. Considering a network of protein-protein interactions and Transcription Factor (TF)-DNA interactions as an electric circuit, conductance of each edge $(u, v)$ is set based on the correlation of \textit{expression values} of $u$ and $v$ with the \textit{target gene}. After solving the electric circuit using Kirchhoff's and Ohm's Laws to get the current through each node and edge, the \textit{causal gene} is taken as the gene with highest value of current flow. \textit{Pathway} is the shortest route between a \textit{target gene} and \textit{causal gene} with the highest total sum of currents across its interactions such that each edge corresponds to an interaction. All such paths together give the \
textit{pathways} for the entire network.

Based on the above approaches, Y. A. Kim et al. suggested an electric circuit based approach with multiple sources and sinks \cite{ref:yoo}, \cite{ref:yoo2}. Conductance is defined  as the average of the absolute value of the Pearson correlation coefficient between \textit{gene expression values} of \textit{target gene} and genes at the endpoints of each link. A system of linear equations based on Ohm's law and Kirchhoff's law is solved to find the voltages of links and thereby to calculate the current value. \textit{Causal genes} are taken as the genes with significant amount of current flow. \textit{Pathway} is the shortest paths in the collection of all maximum current paths for each pair of source and sink. Focusing on the faster computation of voltages of nodes suggested in the previous approach, a fourth order iterative method for fast iterative matrix inversion was proposed in \cite{ref:tij}. Following the calculation of voltages of nodes, causal genes and dysregulated pathways are identified as in \
cite{ref:yoo}, \cite{ref:yoo2}. 

A new  approximation algorithm was proposed in \cite{ref:jee}, based on electric circuit approach to fetch \textit{causal genes} and corresponding \textit{pathways}. Collection of all distinct paths in a reduced network obtained by thresholding the edge-weights is computed. Then, randomized rounding is applied to get the path with maximum current flow which is taken as the \textit{dysregulated pathway} for the corresponding \textit{target} and \textit{causal genes}. The members of the \textit{pathways} are taken as \textit{causal genes}.
 
Here, we infer that the reported works in literature for finding the set of disease causing genes could be evaluated in terms of their accuracy, i.e, the closeness of the set found to the actual set and execution time. During the assessment, it has been observed that in most of the cases, execution time rises with accuracy.

\section{Materials and Methods}
\subsection{Selection of target genes and candidate causal genes} \label{stc}
Gene expression data of Breast cancer and Lung cancer of the species Homosapiens (GSE44024 and GSE43459) and Pancreatic cancer of the species Rattus norvegicus (GSE22537) have been utilized for experimentation. Data have been collected from the National Center for Biotechnology Information (NCBI) sponsored Gene Expression Omnibus data repository \cite{ref:jee}. Platform details and sample information of the datasets are given in Table \ref{Table:in}.
   
\begin{table}[H]
\centering
\label{Table:in}
\begin{tabular}{lcclll}
\hline
\multicolumn{1}{c}{\multirow{2}{*}{Disease}} & \multirow{2}{*}{GEO Accession} & \multirow{2}{*}{\begin{tabular}[c]{@{}c@{}}Sample count\\ (case / control)\end{tabular}} & \multicolumn{3}{c}{\multirow{2}{*}{Platform}} \\
\multicolumn{1}{c}{}                         &                                &                                                                                          & \multicolumn{3}{c}{}                          \\ \hline \hline
Breast cancer &GSE44024 & 4 (2 / 2) & \multicolumn{3}{l}{\begin{tabular}[c]{@{}l@{}}GPL571\\(Affymetrix\\ Human Genome\\ U133A 2.0 Array)\end{tabular}}      \\ \hline
Lung cancer  & GSE43459 & 6 (3 / 3) & \multicolumn{3}{l}{\begin{tabular}[c]{@{}l@{}}GPL6244\\(Affymetrix\\ Human Gene\\ 1.0 ST Array)\end{tabular}}     \\ \hline
Pancreatic cancer& GSE22537  & 18 (9 / 9) & \multicolumn{3}{l}{\begin{tabular}[c]{@{}l@{}}GPL1355\\ (Affymetrix\\ Rat Genome\\ 230 2.0 Array)\end{tabular}}       \\ \hline
\end{tabular}
\caption{Dataset details}
\end{table}

The following steps have been done for each dataset separately. The initial data have been normalized using Robust Multi-array Average method to remove any noise due to non-biological factors \cite{ref:iri}. Then, the genes having statistical significance have been selected using $t$-test with equal variance and $2$-tailed followed by the calculation of $p$-value. Then, significant $q$-values have been computed and genes with $q$-value $<$ 0.05 have been selected. This set of \textit{differentially expressed genes} has been taken as the \textit{candidate causal genes} for Breast cancer and Pancreatic cancer. Considering the data size, first 100 genes after sorting $p$-value in ascending order following the filtering based on $q$-value, have been selected as \textit{candidate causal genes} for Lung cancer. Gene interaction network of molecular interactions has been downloaded from BioGRID database for each \textit{candidate causal gene}. The fetched network data have been filtered to select only the genes 
that are \textit{differentially expressed}. Genes that are linked with transcription factors in the network have been considered as \textit{target genes}. The set of \textit{target genes} has been considered as the set source nodes. The set of genes apart from the \textit{target genes} has been considered as the set of sink nodes. 
\subsection{Selection of benchmark data} \label{sbd}
Data from NCBI sponsored Gene database, Aceview and Uniprot database and Cosmic database have been curated as benchmark data for Breast cancer and Lung cancer of the species, Homosapiens. For this, the set of associated genes corresponding to a particular disease case and species has been fetched from these databases. These fetched genes after removing duplicates have been taken as the benchmark data for each disease. Similarly, data from NCBI sponsored Gene database, Aceview, Uniprot database and literatures from NCBI, Nature, Nucleic Acid Research have been compiled as benchmark data for the species Rattus norvegicus  \cite{ref:donn}, \cite{ref:for}.
\subsection{Formulation of the algorithm}
\subsubsection{Biological Background}\label{bbp}
Each gene $g$ is expressed to a particular level during the process of producing the ultimate products called proteins and this level can be measured as a numerical quantity, known as expression value, $e(g)$ \cite{ref:dud}, \cite{ref:ern}, \cite{ref:mil}, \cite{ref:jos}. Also, each gene interacts with several genes resulting in certain phenotypes \cite{ref:mik}. With the advances in the area of computational biology, molecular interactions can be represented as a network by designating genes/proteins as nodes and edges as associations between end nodes \cite{ref:gab}, \cite{ref:rya}. These associations have different aspects like physical interactions, membership in the same pathway, co-expression and literature co-occurrence \cite{ref:chr}, \cite{ref:dam}. It reveals the fact that two nodes may be linked together in different cases, resulting in a network with multiple edges \cite{ref:al}, \cite{ref:fer}, \cite{ref:pav}. Also, molecular interaction network consists of certain directional links such as 
protein-protein interactions (bidirectional), TF-DNA interactions (directional) and Phosphorylation events (directional) \cite{ref:yoo}, \cite{ref:yoo2}, \cite{ref:yak}, \cite{ref:jee}.

Expression levels or values of genes help to make distinction between healthy and disease cases  in view of the fact that there is an increase or decrease in expression values of some genes in many diseases from that of healthy individuals \cite{ref:omar}, \cite{ref:dev}, \cite{ref:wu}. When a particular gene's expression value changes between two groups of healthy and affected individuals, then the gene is said to be \textit{differentially expressed} \cite{ref:dev}, \cite{ref:jos}, \cite{ref:bar}. Differentially expressed genes may lead to a disease state or to a beneficial state.
\subsubsection{Definitions}
The set of genes that gives rise to a particular disease state is termed as \textit{causal genes}. The set of probable genes of a certain disease is termed as \textit{candidate causal genes}. The set of \textit{candidate causal genes} that are bound by Transcription Factors is referred to as \textit{target genes} \cite{ref:yoo}, \cite{ref:yoo2}, \cite{ref:cho}.\\

As described in Section \ref{bbp}, \textit{candidate causal genes} are connected together in a gene/protein network due to different aspects.
A disease state may be developed by the interference of a \textit{target gene} and a \textit{candidate causal gene} in the normal biological functions of a cell and this relationship between a \textit{target gene} and a \textit{candidate causal gene} is known as dependence \cite{ref:wu}. The dependence between a target gene, $t$ and a candidate causal gene, $c$ is computed as 
\begin{equation}
 \delta(t, c) = P(e(t), e(c))
\end{equation}
where, $P(e(t), e(c))$ is the \textit{Pearson's correlation coefficient} of the \textit{expression values} of genes $t$ and $c$.

Genes with low values of $\delta$ and high values of $\delta$ are crucial in molecular interactions \cite{ref:aga}; so we consider the absolute value of $\delta$ \cite{ref:zhi}. Similarly, in \cite{ref:yoo} and \cite{ref:yoo2}, the edge weight for the edge in the molecular interaction network is taken as the average of the absolute value of the \textit{dependence} of genes (end nodes of the edge) with the \textit{target gene}. The edge weights can be used as a measure of inferring \textit{causal genes} and corresponding \textit{pathways}.
\subsection{Formal notations and definitions}\label{fn}
Let $G_t$ be the set of \textit{target genes}, $G_{cc}$ be the set of \textit{candidate causal genes} and $G_c$ be the set of \textit{causal genes}.

The basic problem is to find $G_c$, the set of \textit{causal genes}, where $G_c \subseteq G_{cc}$ and $G_{cc}$ and $G_t$ are known.

The fundamental decision problem here is 

\quad \quad \quad \quad \quad \quad \quad \quad \quad \quad Does $g_{cc} \in G_c$?

where $g_{cc} \in G_{cc}$ [or, is a candidate causal gene, a causal gene].    

Let us recall that $g_{cc} \in G_c$ if it has a role in disease, and would be determined by the interactions it has with the \textit{target genes}, $g_t \in G_t$, and other \textit{candidate genes} which have a role in the disease.

For each \textit{target gene} $g_t$, a weighted graph (network) is defined with $g_t$ as the source, members in $G_{cc}$ as the other (non-source) nodes, and molecular interactions as the edges. Edge weights reflect the role of the genes represented by the end nodes in causing the disease. It may be realized at this point that there would be $|G_t|$ such graphs, each containing $|G_{cc}|+1$ nodes. Let $(\mathcal{G}_t, w_t)$ be such a network for \textit{target gene}, $g_t$.

\quad \quad \quad \quad \quad \quad \quad \quad $\mathcal{G}_{t}.V = \{g_t\} \cup G_{cc}$\\
$\mathcal{G}_{t}.E$ is defined by molecular interactions, as described next. Let $c(g_{cc}, g_t) = \delta(g_t, g_{cc})$.

Nodes  in $\mathcal{G}_{t}.V$ would have had multiple edges between them, as in Section \ref{bbp}. The molecular interaction network is simplified, to get ordinary graphs, using the following method. 

The weight function $w_t : \mathcal{G}_t.V \times \mathcal{G}_t.V \rightarrow [0, +1]$ is defined as,
\begin{equation}
 w_t(g_x, g_y) = \frac{|c(g_x, g_t)| + |c(g_y, g_t)|}{2}
\end{equation}
where, $g_x$, $g_y$ are \textit{candidate causal genes} and $g_x \sim g_y$ (i.e., they have an interaction in any of the three ways outlined in Section \ref{bbp}) \cite{ref:yoo}, \cite{ref:yoo2}.

Thus, multiple edges defined in Section \ref{bbp} get substituted with a single weighted edge in $\mathcal{G}_t$.

The genes at the endpoints of edges with higher edge weights \cite{ref:yoo}, \cite{ref:yoo2} are said to have a role in the disease, or are \textit{causal genes}.

The weight of a path $p$ from $g_t$ to $g_{cc} = \sum_{e \in g_t \stackrel{p}{ \ensuremath{\leadsto} } g_{cc}} w_t(e).$ 

The paths from $g_t$ to \textit{candidate causal genes} having high weight values are called \textit{dysregulated pathways}. There may be multiple \textit{dysregulated pathways} in any $(\mathcal{G}_t, w_t)$. It may be recalled that there are $|G_t|$ such $(\mathcal{G}_t, w_t)$, or interaction networks, and each typically contains multiple \textit{dysregulated pathways}.

Genes corresponding to nodes belonging to the \textit{dysregulated pathways} are called the \textit{causal genes}. i.e., $g_{cc} \in G_c$, if there exists $g_t$ and a path $p$ such that $g_t \stackrel{p}{\leadsto} v$ and $g_{cc} \in p$ where, $v$ is a \textit{candidate causal gene}, $g_t$ is a \textit{target gene} and $g_t \stackrel{p}{\leadsto} v$ is a \textit{dysregulated pathway} in $(\mathcal{G}_t, w_t)$. 

In the above description, the terms \textquotedblleft high\textquotedblright \hspace{0.5 mm} edge weights and \textquotedblleft high\textquotedblright \hspace{0.5 mm} weighted paths have been used. As these adjectives are abstract, they need to be quantified.
\subsection{Towards a new algorithm} \label{tna}
The problem defined in Section \ref{fn} is explained and stated below, without abstract adjectives.

A protein network can be represented as an edge-weighted directed graph with each $g_{t}$ as the source vertex and $w_t(u,v)$ as the edge-weight between nodes $u$ and $v$ defined previously. Nodes represent genes or proteins and edges represent molecular interactions.

$(\mathcal{G}_t, w_t)$ obtained for each $g_t \in G_t$ is further reduced, by eliminating \textquotedblleft low\textquotedblright \hspace{0.5 mm} weighted edges.

\textquotedblleft High\textquotedblright \hspace{0.5 mm} and \textquotedblleft low\textquotedblright \hspace{0.5 mm} are decided based on a \textit{threshold}, which is defined as follows.

In $(\mathcal{G}_t, w_t)$, the \textit{threshold} $\tau$ is defined as
\begin{equation} \label{ta}
 \tau = \frac {{\sum_{e \in \mathcal{G}_t.E}} w_t(e)}{|\mathcal{G}_t.E|}
\end{equation}
The set of edges to be removed, $R$, is computed as follows.
\begin{equation}
 R = \{e | e \in \mathcal{G}_t.E \wedge w_t(e) < \tau\}
\end{equation}
The edges in $R$ are then removed from $(\mathcal{G}_t, w_t)$ \cite{ref:jee}.

The paths having high path weights (as defined earlier) from $g_t$ to all \textit{candidate causal genes}, in graphs for all $g_t \in G_t$ are the \textit{dysregulated pathways} and all vertices belonging to \textit{dysregulated pathways} are \textit{causal genes}.


Given $n$ graphs $(\mathcal{G}_i, w_i)$ with source vertex $i \in G_t$ and $\mathcal{G}_i.V = i \cup G_{cc}$. Find $G_c$.

\begin{multline}
\label{eq:gc1}
 G_c = \cup_{\forall (\mathcal{G}_i, w_i)} \{g_{cc} | w_i(g_{cc}, g_i) > \tau \\
\wedge \exists p : g_i \stackrel{p}{\leadsto} g_{cc} \\
\wedge \forall e \in p \quad w_i(e) > \tau \} 
\end{multline}

where, $\tau$ is the \textit{threshold} value defined in Equation \ref{ta}.
\\
Find \textit{dysregulated pathways} $D_p$ as all the paths $p$ satisfying the above condition. i.e.,
\begin{equation}
\label{eq:dp1}
 D_p = \{p | \forall e \in p \quad w_i(e) > \tau\}
\end{equation}

It is inferred that every research work described in Section \ref{intro} attempts to find the complete set $G_c$ and all \textit{dysregulated pathways}. However, for realistic biological networks, the size of the sets and the number of paths is very large for practical computation \cite{ref:yoo}, \cite{ref:yoo2}. In order to get the results in reasonable computation time, the following heuristic is used in this work. \textquotedblleft In a biological network, higher the \textit{degree} of a node, more relevant it is\textquotedblright \hspace{0.5 mm} \cite{ref:jon}, \cite{ref:goh}, where \textit{degree} of a node is the number of its neighboring edges. \textit{Degree} of a node, $v$ is denoted by $deg(v)$.

Hence, each time a path $p \in D_p$ is explored in $(\mathcal{G}_i, w_i)$, a vertex $g_{cc}$ with $deg(g_{cc}) = \delta(\mathcal{G}_i, w_i)$ is removed. $\delta(\mathcal{G}_i, w_i)$ is the minimum degree of $(\mathcal{G}_i, w_i)$. Let $v_\delta$ be the node with minimum degree in $(\mathcal{G}_i, w_i)$. Then, $(\mathcal{G}_i, w_i)$ is updated such that $\mathcal{G}_i.V = \mathcal{G}_i.V \setminus v_\delta$. Next, a path, $p \in D_p$ is traversed in $(\mathcal{G}_i, w_i)$ and repeat this process till $|V|=  k$, where $k$ is an integer.

The process of removing $v_\delta$ can be made efficient in terms of execution time by mapping $G_{cc}$ along with the degree of each node, $deg(g_{cc})$ to \textit{min-heap} data structure. \textit{Min-heap} is the underlying data structure of priority queues with the property of $A[parent(u)] \leq A[u]$, where $u$ is a node other than root node and $parent(u)$ is the parent node of $u$ \cite{ref:cor}. Here, priority is given in terms of the decreasing order of degree. Highest priority is assigned to the vertex $v_\delta$ to ensure that the vertex with the least degree is removed. Considering each node and its degree in the molecular interaction network as the elements of the \textit{min-heap}, $v_\delta$ is deleted by removing the root node of the \textit{min-heap}. Then, this data structure is reorganized to bring the highest priority node or $v_\delta$ at the root \cite{ref:cor} and the process specified earlier continues till the number of nodes $= k$.

Each time $v_\delta$ is removed, its adjacent edges are also removed, thereby reducing the complex nature of biological networks which in turn results in identification of relevant genes and corresponding pathways within practical time interval.

Now, definitions of $G_c$ and $D_p$ are updated as follows:

\begin{multline}
 G_c = \cup_{\forall (\mathcal{G}_i, w_i)} \{g_{cc} | w_i(g_{cc}, g_i) > \tau \\
\wedge \exists p : g_i \stackrel{p}{\leadsto} g_{cc} 
\wedge \forall e \in p \quad w_i(e) > \tau \\
\wedge \exists g_{cc} : deg(g_{cc}) > \delta(\mathcal{G}_i, w_i)
\} 
\end{multline}

Here, an additional characteristic of \textit{causal genes} is incorporated apart from the attributes described in Equation \ref{eq:gc1}: \textit{causal genes} tend to be higher \textit{degree} nodes.

\textit{Dysregulated pathways} $D_p$ as all the paths $p$ satisfying the mentioned conditions. i.e.,
\begin{equation}
 D_p = \{p | \forall e \in p \quad w_i(e) > \tau \wedge \exists v \in p : deg(v) > deg(v_\delta) \}
\end{equation}

Finally, motivated by the approach in \cite{ref:jee}, \textit{randomized rounding}, an efficient approach to design an \textit{approximation algorithm} is done to fetch the most relevant paths in terms of weight, $w_i(e)$ and degree, $deg(g_{cc})$ in $(\mathcal{G}_i, w_i)$. The randomized rounding approach resulted in a factor $\frac{\psi}{\eta \kappa}$ algorithm. Here, $\psi$ is the total number of paths in $\Psi$, where, $\Psi$ is the collection of all distinct paths in the reduced network. $\eta$ is the total number of distinct edges in $\Psi$ and $\kappa$ is the number of occurrences of a randomly selected edge in $\Psi$.  

Its underlying principle follows. $\sum_{\forall e \in p} w_i(e)$ is calculated for $(\mathcal{G}_i, w_i)$, where $p \in D_p$ . Let $\mu_p$ be $max(\sum_{\forall e \in p} w_i(e))$ in $(\mathcal{G}_i, w_i)$. \textit{Dysregulated pathway} is taken as the path having value $\mu_p$ and the members of the pathway is taken as the \textit{causal genes} for $(\mathcal{G}_i, w_i)$. Therefore, definitions of $G_c$ and $D_p$ are reformed by adding the following  attribute: \textit{causal genes} and thereby \textit{dysregulated pathways} are constituted of maximum weighted paths after \textit{randomized rounding}. Final definitions of $G_c$ and $D_p$ after joining all the mentioned attributes together are as given below. 

\begin{multline}
 G_c = \cup_{\forall (\mathcal{G}_i, w_i)} \{g_{cc} | w_i(g_{cc}, g_i) > \tau \\
\wedge \exists p : g_i \stackrel{p}{\leadsto} g_{cc} 
\wedge \forall e \in p \quad w_i(e) > \tau \\
\wedge \exists g_{cc} : deg(g_{cc}) > \delta(\mathcal{G}_i, w_i) \\
\wedge \exists p : \forall e \in p \quad \Sigma w_i(e) = \mu_p
\} 
\end{multline}

\textit{Dysregulated pathways} $D_p$ as all the paths $p$ satisfying the stated conditions. i.e.,
\begin{multline}
 D_p = \{p | \forall e \in p \quad w_i(e) > \tau \wedge \exists v \in p : deg(v) > deg(v_\delta) \\ \wedge \forall e \in p \quad \Sigma w_i(e) = \mu_p \}
\end{multline}

\subsection{Implementation and Validation of the results}
Comparative study of the related works in literature and the proposed method has been performed based on the accuracy of results and the execution time. All these methods have been implemented in Intel \textregistered Core i7-3770 CPU @ 3.40GHz X 8 machine with memory (RAM) capacity of 16 GB by using the programming languages R and C. Also, the effectiveness of the Algorithm \ref{alg:approx} has been explored by using Reactome Pathway Database to provide its biological significance. 
\subsubsection{Analysis based on $\tau $ and $k$} \label{tk}
As the threshold defined in \ref{ta}, is quite critical for removing the low-weighted edges, its effect on identifying causal genes and dysregulated pathways has been simulated. For this, apart from implementing Algorithm \ref{alg:approx}, an algorithm without using steps 4, 5 and 6 in Algorithm \ref{alg:approx} has been implemented. Also, Algorithm \ref{alg:approx} has been implemented against different values of $k$ which is defined in Section \ref{tna} for the datasets.  

\section{Discussion and Results}
\subsection{Approximation algorithm with graph reduction}\label{na}
Pseudo code for the algorithm to identify causal genes and dysregulated pathways is given in Algorithm \ref{alg:approx}. The algorithm considers a molecular interaction network with multiple sources and sinks. Source set is the set of target genes and sink set is the set of candidate causal genes other than target genes. Proof of the approximation factor is given in Appendix \ref{app}.
			\begin{algorithm}[!hbt]
				\caption{Infer causal genes and dysregulated pathways}
				\label{alg:approx}
				\textbf{Input:} $n$ graphs $(\mathcal{G}_i, w_i)$ with source vertex $i \in G_t$ and sink vertex $g_{cc} \in G_{cc}$\\
				\textbf{Output:} $G_c$ and $D_p$\\
				\line(1, 0){250}\\
				\begin{algorithmic}[1]
                                       \ForAll{$i \in G_t$} \label{step_init}
                                         \ForAll{$g_{cc} \in G_{cc}$}
                                            \State $s \gets g_{cc}$
					\State Calculate $\tau$
					\State Find $R$
					\State Remove $R$ from $(\mathcal{G}_i, w_i)$
                                        \Repeat 
                                            \State $\Psi$ = $\Psi \cup Findpath((\mathcal{G}_i, w_i), i)$
                                        \Until{($|V| = k$)} 
					\State $\psi \gets |\Psi|$
					\If{$\psi = 0$}
						\State goto step \ref{step_init}
					\ElsIf{$\psi = 1$}
                                                \State $D_p = \Psi$
						\State Find $G_c$
              					\State goto step \ref{step_init}
					\Else		
					
					\State Find $\eta$, total number of distinct edges in $\Psi$
					\State Randomly pick an edge, $e$ from a path in $\Psi$
					\State Calculate $\kappa$, number of occurrences of \textit{e} in $\Psi$
                                        \State $r \gets \frac{\eta\kappa}{\psi}$
					\For{$i = 1$ \textbf{to} $r$}
						\State Randomly select a path $p$ in $\Psi$
						\State Calculate $\sum_{\forall e \in p} w_i(e)$
					\EndFor
					\State $D_p$ $\gets$ Path with value $\mu_p$
					\State Find $G_c$
                                        \EndIf \label{step_end}
					\EndFor
                                        \EndFor
				\end{algorithmic}
			\end{algorithm}

\begin{algorithm}[!hbt]
\begin{algorithmic}
  \Procedure{$Findpath$}{$(\mathcal{G}_i, w_i), v$}
 
   \If{$v = s$}
      \State $p = p \cup v$
      \State $\Psi \gets p$
      \State $\mathcal{G}_1.V = \mathcal{G}_i.V \setminus \{i, s\}$
      \State Calculate $deg(g_{cc})$ for all $g_{cc} \in \mathcal{G}_1.V$						
      \State Assign priority based on decreasing order of degree
      \State Make a $min-heap$ for  $(\mathcal{G}_1)$
      \State Find $v_\delta$
      \State Delete $v_\delta$ from $min-heap$
      \State Update $\mathcal{G}_i.V = \mathcal{G}_i.V \setminus v_\delta$
  \Else
    \ForAll{neighbours $adj$ of $v$} 
      \If{$adj$ is unexplored}
          \State $p = p \cup v$
          \State $Findpath((\mathcal{G}_i, w_i), adj)$
      \EndIf
    \EndFor      
  \EndIf
  \EndProcedure
\end{algorithmic}
\end{algorithm}
\subsection{Implementation and Validation of the results}\label{im}
All the related works in literature and the proposed algorithm have been implemented as specified in Section \ref{stc} on multiple datasets. For Breast cancer dataset, $G_t$ consists of 41 genes and $G_{cc}$ consists of 309 genes, for Lung cancer dataset, $G_t$ consists of 7 genes and $G_{cc}$ consists of 70 genes and for Pancreatic cancer dataset, $G_t$ consists of 46 genes and $G_{cc}$ consists of 140 genes. $(\mathcal{G}_t, w_t)$ have 8403 edges and 350 nodes with a total of 41 such graphs for Breast cancer dataset, 466 edges and 77 nodes with a total of 7 such graphs for Lung cancer dataset and 729 edges and 186 nodes with a total of 46 such graphs for Pancreatic cancer dataset. Benchmark data as specified in Section \ref{sbd} have been obtained for the cases Breast cancer, Lung cancer and Pancreatic cancer as 26 genes, 21 genes and 83 genes respectively. 

Let us denote Breast cancer dataset as dataset I, Lung cancer dataset as dataset II and Pancreatic cancer dataset as dataset III.
\subsubsection{Simulation study based on $\tau$}
The effect of $\tau$ on identifying causal genes and dysregulated pathways has been simulated by considering multiple graphs with $SLCO2A1$ as source for a subset of sink nodes on dataset I, $Cdk6$ as source for a subset of sink nodes on dataset II and $Acaca$ as source for a subset of sink nodes on dataset III, as specified in Section \ref{tk}.  The results are summarized and tabulated in Table \ref{Table:t1}.
\begin{table}[H]
 \centering
{\begin{tabular}{@{}ccc@{}}\hline Dataset & \# Causal Genes & Execution Time \\\hline\hline
I & [86][87]  & [435.35 Sec.][737.33 Sec.] \\
II & [24][35] & [1.503 Sec.][1.893 Sec.] \\
III & [8][9]  & [45.5 Sec.][800.119 Sec.] \\\hline
\end{tabular}}
\caption{Effect of $\tau$ with the order [with $\tau$][without $\tau$] \label{Table:t1}} 
\end{table}

There is a slight increase in the number of causal genes identified in the method without using $\tau$; but, it compromises on execution time while considering all the genes.
\subsubsection{Analysis based on $k$}
Algorithm \ref{alg:approx} has been implemented against different values of $k$ as defined in Section \ref{tna} for the datasets. Considering the size of the datasets, for dataset I, a subgraph of $(\mathcal{G}_t, w_t)$ has been considered by selecting first 100 genes after sorting $p$-value, resulting in 8 graphs of 432 edges and 74 nodes with benchmark data consisting of 10 genes.  The receiver-operating characteristic (ROC) analysis has been used to compare the performance of the algorithm under various values of $k$. The values of $k$ are selected as 10\% of $|V|$, 20\% of $|V|$, 30\% of $|V|$ and 40\% of $|V|$. The ROC curve plots the true positive rate (TPR) / sensitivity versus the false positive rate (FPR) / (100-specificity) where, TPR is the percentage of causal genes which are correctly identified based on benchmark dataset and FPR is the percentage of causal genes which are not present in the benchmark dataset. To compare different curves obtained by ROC analysis, the area under curve (AUC) for 
each curve has been calculated, which is given in Table \ref{table:AUC}. Higher the value of AUC, better the result is. Also, the execution time for each case is taken and it is summarized in Table \ref{timeIandII}. When the ROC curve and execution time are compared, k=10\% of $|V|$ has shown better performance and it has been selected as the cut-off value in Algorithm \ref{alg:approx} provided overall accuracy is same in all cases. 
\begin{table}[H]
 \centering
 {\begin{tabular}{@{}cccc@{}}\hline Values of k & Dataset I & Dataset II & Dataset III\\\hline\hline
10\% of $|V|$  &  0.0294 & 0.6932 & 0.656\\
20\% of $|V|$ & 0.0294 & 0.5263 & 0.694\\
30\% of $|V|$&  0.0294 & 0.3982 & 0.579\\
40\% of $|V|$&  0.0294 & 0.4463 & 0.552\\\hline
\end{tabular}}
\caption{AUC for different values of k\label{table:AUC}}
\end{table}

\begin{table}[H]
 \centering
{\begin{tabular}{@{}cccc@{}}\hline Values of k & Dataset I & Dataset II & Dataset III \\\hline\hline
10\% of $|V|$  & 0.16 & 6.68 & 90.291\\
20\% of $|V|$ & 1.5 & 9.57 & 148.28\\
30\% of $|V|$&  10.9 & 10.38 & 167.206\\
40\% of $|V|$&  14.06 & 14.24 & 1637\\\hline
\end{tabular}}
\caption{Execution time (Seconds) for different values of k \label{timeIandII}} 
\end{table}

\subsection{Measuring the accuracy of results and Analysis based on the execution time}
\begin{figure}
\includegraphics[scale = 0.5]{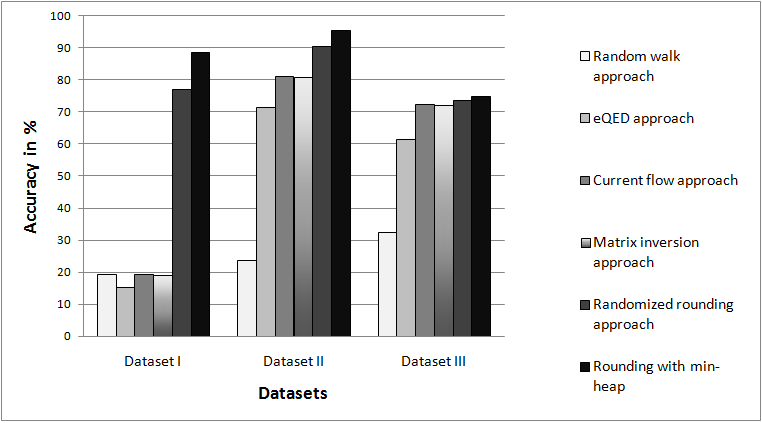}
\caption{Comparison based on the accuracy of results}
\label{accTime}
\end{figure}

Accuracy is defined as the measurement of closeness between the benchmark data and the actual results. It is calculated as the percentage of causal genes which are correctly identified based on benchmark dataset. Comparison based on the accuracy of results is given in Figure \ref{accTime} and the execution time is given in Table \ref{et}.\\
\begin{table}[H]
 \centering
  {\begin{tabular}{lcccc}\hline Approaches & Dataset I & Dataset II & Dataset III \\\hline\hline
Random walk approach &  65 (Min.) & 5 (Sec.) & 1 (Min.)\\
eQED approach &  33 (Min.) & 4 (Sec.) & 0.8 (Min.)\\
Current flow approach &  37 (Min.) & 1801 (Hours) & 1226 (Min.)\\
Matrix inversion approach &  34 (Min.) & 1789 (Hours) & 1212 (Min.)\\
Randomized rounding approach&  4114 (Min.) & 53 (Hours) & 1.73 (Min.)\\
Rounding with min-heap&  2155 (Min.) & 8.61 (Sec.) & 1.12 (Min.)\\\hline
\end{tabular}}
\caption{Comparison based on the execution time}
\label{et}
\end{table}
Observations made based on these results are given next.
In random walk approach, the Pearson's correlation coefficient of the \textit{gene expression levels} of genes at each node and the \textit{target gene} is calculated and transition probability has obtained using this value. Random walk based on transition probabilities is done multiple times. \textit{Disease causing genes} including their \textit{pathways} are obtained as the result. Random walk approach takes less amount of execution time, but the number of genes identified and the accuracy of results are less. The molecular interaction is considered as an electric circuit according to other approaches in literature. The conductance of each link is calculated based on the Pearson's correlation coefficient of the \textit{gene expression levels} of genes at each node and \textit{target gene}. The  Kirchhoff's current law and Ohm's law are utilized to obtain voltages. The current is calculated using Ohm's law for each edge in the network. \textit{Disease causing genes} and \textit{dysregulated pathways} are 
identified using each of these methods. eQED approach takes least amount of execution time, but gives lesser number of genes and low accuracy compared to methods other than random walk approach. Electric circuit approach with multiple sources and sinks and fast iterative matrix inversion return same number of disease causing genes with same accuracy. Here, genes receiving current of at least 70\% of the maximum current among all genes are considered in each iteration and thereby its execution time is greater in most of the cases. Since, fast iterative matrix inversion uses fourth order iterative method, execution time is less compared to electric circuit approach with multiple sources and sinks. Approximation algorithm outperforms other methods in terms of the number of disease causing genes identified and the accuracy of results. But, it takes much more time compared to random walk approach and eQED approach. Here, before applying randomized rounding, all possible paths are identified which in turn results 
in increased execution time. 

The proposed algorithm, Approximation algorithm with graph reduction has been implemented on multiple datasets and the sets of \textit{causal genes} and associated \textit{pathways} for the disease cases have been identified. Initially, a set of less-weighted edges are removed. Then, a set of vertices are removed together with the exploration of a set of distinct simple paths by making sure that a vertex of least degree is deleted. Min-heap data structure is utilized for time efficient vertex removal. This process repeats $k$ times with its value as 10\% of $|V|$. Finally, most relevant paths are identified using randomized rounding. This newly proposed approximation algorithm with graph reduction defeats all other methods by identifying more number of genes within less time interval.

The sets of all newly identified \textit{disease causing genes} apart from the known genes in the benchmark datasets as well as the resultant sets of identified \textit{causal genes} and \textit{dysregulated pathways} for all the datasets are given in the Supplementary material.
\section{Conclusion and Future scope}
Computationally intensive process of identifying causal genes and related pathways from the huge molecular interaction networks has been addressed in this paper. The huge size of the molecular interaction network is reduced by retaining more relevant nodes and edges in terms of edge-weights and connectivity. Then, most appropriate nodes, thereby paths in terms of these parameters are selected by using the concept of \textit{approximation}. Experimentations proved that the newly proposed approximation algorithm with graph reduction outperforms all other methods by identifying more number of genes within lesser time. 
\bibliographystyle{IEEEtran}
\bibliography{Document}
\end{document}